\UseRawInputEncoding
\documentclass[a4paper,11pt]{article}

\usepackage{amsthm,amsmath}
\usepackage{algorithm,algpseudocode}
\usepackage[%
bookmarks=true,
ocgcolorlinks,%
linkcolor=blue,%
filecolor=blue,%
citecolor=blue,%
urlcolor=blue]{hyperref}
\usepackage{amssymb}
\usepackage{mathtools}
\usepackage{graphicx}
\usepackage{psfrag}
\usepackage{verbatim}
\usepackage{tikz}
\usetikzlibrary{quantikz2}

\usepackage[font={footnotesize}]{caption}

\newcommand{\comments}[1]{}

\numberwithin{equation}{section}

\title{Encoding of Probability Distributions for Quantum Monte Carlo Using Tensor Networks}

\begin{document}

\maketitle

\begin{center}
    \begin{tabular}{c}
        Ant\'onio Pereira\textsuperscript{1,*}, Alba Villarino\textsuperscript{1,*}, Aser Cortines\textsuperscript{2,*}, \\
        Samuel Mugel\textsuperscript{3}, Rom\'an Or\'us\textsuperscript{1,4,5}
    \end{tabular}
    \\
    \vspace{0.5em}
    \small{
    \begin{tabular}{c}
        \textit{\textsuperscript{1}Multiverse Computing, Paseo de Miram\'on 170, 20014 San Sebasti\'an, Spain} \\
        \textit{\textsuperscript{2}Multiverse Computing, rue de la Croix Martre, 91120 Palaiseau, France} \\
        \textit{\textsuperscript{3}Multiverse Computing, 192 Spadina Ave, 509 Toronto, Canada} \\
        \textit{\textsuperscript{4}DIPC, Paseo Manuel de Lardizabal 4, E-20018 San Sebasti\'an, Spain} \\
	\textit{\textsuperscript{5}Ikerbasque Foundation for Science, Maria Diaz de Haro 3, E-48013 Bilbao, Spain}\\
        \textsuperscript{*}\textit{\{antonio.pereira, alba.villarino, aser.cortines\}@multiversecomputing.com}
    \end{tabular}
    }
\end{center}

\vspace{1em}

\begin{center}
    \begin{tabular}{c}
        Victor Leme Beltran\textsuperscript{4,*}, J.V.S. Scursulim\textsuperscript{4,*}, Samura\'i Brito\textsuperscript{4,*}
    \end{tabular}
    \\
    \vspace{0.5em}
    \small{
    \begin{tabular}{c}
        \textit{\textsuperscript{4}Instituto Ita\'u de Ci\^encia, Tecnologia e Inova\c{c}\~ao} \\
        \textsuperscript{*}\textit{\{victor.beltran, jose.scursulim, samurai.brito\}@itau-unibanco.com.br}
    \end{tabular}
    }
\end{center}
  
\vspace{2.5em}

\begin{abstract}
The application of Tensor Networks (TN) in quantum computing has shown promise, particularly for data loading. However, the assumption that data is readily available often renders the integration of TN techniques into Quantum Monte Carlo (QMC) inefficient, as complete probability distributions would have to be calculated classically. In this paper the tensor-train cross approximation (TT-cross) algorithm is evaluated as a means to address the probability loading problem. We demonstrate the effectiveness of this method on financial distributions, showcasing the TT-cross approach's scalability and accuracy. Our results indicate that the TT-cross method significantly improves circuit depth scalability compared to traditional methods, offering a more efficient pathway for implementing QMC on near-term quantum hardware. The approach also shows high accuracy and scalability in handling high-dimensional financial data, making it a promising solution for quantum finance applications.
\end{abstract}

\newpage

\section{Introduction}
\label{sec:intro}
Quantum computing has emerged as a transformative technology with the potential to revolutionize various fields, including cryptography~\cite{scholten2024assessing}, optimization~\cite{abbas2023quantum} and financial modeling~\cite{Woerner_2019}. In finance, one notable application is the use of Quantum Amplitude Estimation~\cite{Brassard_2002} to accelerate Monte Carlo computations, a technique often referred to as Quantum Monte Carlo (QMC). Monte Carlo (MC) methods are widely used for solving problems involving uncertainty, random processes, or high-dimensional integrals. These methods find applications not only in finance but also in physics, where they model particle interactions, radiation transport, and quantum phenomena. In addition, they are employed in engineering for tasks such as reliability analysis, optimization, and risk assessment. Applications extend further to computer science, where they are employed in machine learning, graphics rendering, and approximate inference, and to biology and medicine, where they model population dynamics and radiation dose calculations. Given their flexibility, MC methods play an essential role in both theoretical and applied research across a broad spectrum of disciplines. Classical MC methods, while powerful, suffer from limitations in computational efficiency, especially when dealing with high dimensional problems. QMC promises a quadratic speed-up over classical MC methods, potentially transforming computational performance in financial modeling.

While classical MC methods are powerful, they are often limited by computational inefficiencies, particularly in high-dimensional problems. Quantum Monte Carlo holds the promise of a quadratic speed-up over its classical counterparts, presenting a opportunity to enhance computational performance. However, this theoretical speed-up has recently been brought into question~\cite{Herbert_2021}. The concern is that the speed-up is typically measured in terms of query complexity rather than overall computational complexity, and these are not necessarily equivalent. Querying a quantum computer entails significant overheads absent in classical computations, such as state preparation and error correction. When these additional operations are considered, the actual computational advantage may be substantially reduced or even negated. 

A significant bottleneck in QMC methods is state preparation, specifically the probability loading problem, which involves translating probability distributions into quantum states. This task is particularly challenging due to its poor scalability and the complexity of its computational steps. The Grover-Rudolph method~\cite{grover2002creating}, commonly used for this purpose, requires a series of computational steps that become increasingly complex as the precision of the state preparation increases. This preparation process is not only time-consuming but also prone to errors, often undermining the claimed advantages of QMC. 

Different approaches can be found in the literature to address this problem, including variational approaches using Quantum Generative Adversarial Networks (qGANs)~\cite{Zoufal_2019} and Quantum Walks \cite{Pachat_2022}. However, these methods face significant challenges that limit their practical application in industrial settings. By training quantum variational circuits to approximate target distributions, qGANs provide a flexible approach for encoding probability distributions. However, qGANs require substantial training times and extensive classical-quantum integration to develop the final circuit. This process can be more complex and time-consuming than the Monte Carlo algorithms it aims to improve. Conversely, approaches based on quantum walks, although effective and relatively straightforward to implement on current quantum machines, are limited to discrete value problems (e.g., Binomial models), failing to cover the full range of potential pricing applications. 

This work introduces and evaluates an innovative solution to the probability distribution loading problem using Tensor Network (TN) techniques \cite{Or_s_2014}. Our approach uses the tensor-train cross approximation algorithm (TT-cross)~\cite{OSELEDETS201070}  to accurately represent complex probability distribution functions in a tensor-train (TT) format, also known as Matrix Product States (MPS) in physics contexts, which can be directly mapped to quantum gate operations~\cite{Ran_2020}. The training and mapping routines are thus entirely performed on classical machines and come with guarantees for convergence and complexity, resulting in scalable circuits. This approach maintains high accuracy while significantly enhancing circuit scalability. We test the effectiveness of our approach on real industrial use-cases provided by Ita\'u Unibanco, the largest bank in Brazil, and demonstrate that our method significantly enhances circuit scalability. 

Our work was developed independently of the study by Sakaue et al.~\cite{sakaue2024learning}, which uses similar concepts in learning and representing noisy functions. In contrast, our approach specifically addresses the distribution loading problem with a focus on enhancing the Grover-Rudolph method. This technique aims to significantly improve the scalability and efficiency of quantum algorithms for financial applications and derivative pricing, particularly focusing on probability distributions that are relevant to the business of Ita\'u Unibanco.

The article is structured as follows. In Section~\ref{sec:tn}, there is an introduction to tensor networks subdivided into two subsections: Tensor learning algorithms, which focuses on the TT-cross approximation, and Quantization, which includes definitions of the sequential, mirroring, and interleaving methods. In Section~\ref{sec:tn_to_qc}, we delve into the mapping of TT to Quantum Circuits (QC). Finally, in Sections~\ref{sec:res} and~\ref{sec:conclusion}, we present the study results and the conclusions, respectively.

\section{Tensor Networks}
\label{sec:tn}

Tensor networks~\cite{Or_s_2014} are a mathematical framework used to efficiently represent and manipulate high-dimensional data. Originating from quantum many-body physics, they have become a powerful tool in various fields, including quantum computing~\cite{patra2024efficient} and machine learning~\cite{novikov2015tensorizing}. 
The success of tensor networks in these domains is primarily due to their ability to exploit the inherent structure of correlations present within many real datasets by encoding them in a low-rank representation. This enables compression of high-dimensional datasets, significantly reducing computational complexity and memory requirements, making tensor networks a key tool for handling large-scale, intricate data structures.

\begin{figure}[htb!]
\centering
\begin{tikzpicture}[
  node distance=2cm,
  label distance=-0.3cm,
  font=\small
]

\node[circle, draw, minimum size=0.8cm] (a) at (1,0) {};
\node[circle, draw, minimum size=0.8cm] (b) at (5,0) {};
\node[circle, draw, minimum size=0.8cm] (c) at (1,-2) {};
\node[circle, draw, minimum size=0.8cm] (d) at (5,-2) {};

\node at (-0.6,0.6) {(a)};
\node at (3.6,0.6) {(b)};
\node at (-0.6,-1.4) {(c)};
\node at (3.6,-1.4) {(d)};
\node [right=0.5cm of b] {vector};
\node [right=0.5cm of d] {rank-3 tensor};
\node [right=0.5cm of a] {scalar};
\node [right=0.5cm of c] {matrix};

\draw (b) -- ++(-1,0);
\draw (c) -- ++(1,0);
\draw (c) -- ++(-1,0);
\draw (d) -- ++(-1,0);
\draw (d) -- ++(1,0);
\draw (d) -- ++(0,-1);

\end{tikzpicture}
\caption{Linear algebra objects as Tensor Network diagrams.}
\label{fig:tensors}
\end{figure}
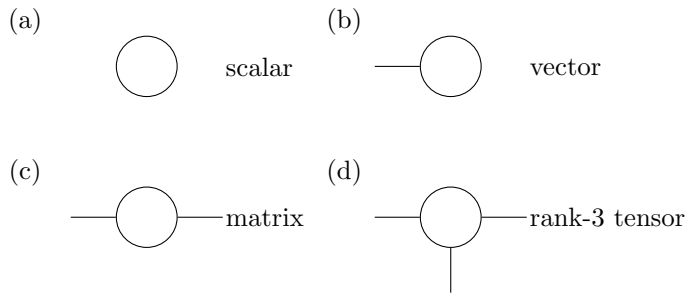

Tensors are multi-dimensional arrays of numbers, that generalize linear algebra objects such as scalars, vectors, and matrices. For instance, a scalar is a zero-dimensional tensor, a vector is a one-dimensional tensor, a matrix is a two-dimensional tensor, and so on. A basic diagrammatic representation of this can be found in Figure~\ref{fig:tensors}. A tensor network represents a high-dimensional tensor as a network of interconnected low-dimensional tensors. Each node in the network represents a tensor, and the edges connecting the nodes represent the indices over which summations (contractions) are performed. This structure allows for the efficient computation of operations involving high-dimensional tensors by breaking them down into simpler, more manageable pieces.

\subsection{Tensor Learning algorithms}
\label{subsec:tt-cross}
Tensor learning algorithms efficiently capture and represent probability distributions using tensor networks. By decomposing complex data into simpler, lower-dimensional components, tensor learning algorithms significantly reduce computational resources and storage requirements. They are widely used in quantum computing, machine learning, and data compression, providing a scalable approach to handle vast and intricate datasets that would otherwise be computationally prohibitive.

One such algorithm is the TT-cross approximation. This generalizes the matrix cross approximation to tensors using the structure of tensor trains (TT) to mitigate the curse of dimensionality~\cite{Alcazar_2022,Ran_2020,rudolph2022decomposition}. 

A tensor \( A \) of order \( d \) (a \( d \)-dimensional array) can be approximated with a TT-format as follows:

\begin{equation}\label{eq:tt}
A(i_1, i_2, \ldots, i_d) \approx \sum_{r_0, r_1, \ldots, r_d} G_1(r_0,i_1,r_1) G_2(r_1,i_2,r_2) \cdots G_d(r_{d-1},i_d,r_d),
\end{equation}

\noindent
where each \( G_k \) for \( k = 1, \ldots, d \) is a three-dimensional tensor, called a TT core. The mode sizes are \( n_k \) for \( k = 1, \ldots, d \), and the ranks are \( r_k \) for \( k = 0, \ldots, d \), with \( r_0 = r_d = 1 \) to maintain dimensional consistency.

The TT-cross method approximates the entire tensor by identifying a subset structured similarly to a cross in matrices.

The algorithm proceeds as follows:
\begin{enumerate}
    \item  Identifying Index Sets: Select right index sets \( J_k \) and left index sets \( I_k \) that indicate tensor positions for sampling. These sets form left-nested and right-nested sequences, respectively, which target submatrices with near-maximal volume.
    \item Computing Tensor Cores: Compute a sequence of tensor cores \( G_k \) by interpolating the tensor over the selected entries. This process requires solving smaller optimization problems, often using Singular Value Decomposition (SVD) or QR decomposition, to identify the optimal low-rank approximation at each stage.
    \item Assembling TT Approximation: Construct the final TT approximation from the computed cores \( G_k \), ensuring the approximation adheres to the sampled entries according to \( I_k \) and \( J_k \).
\end{enumerate}

This iterative process refines the tensor cores \( G_k \), and the index sets \( I_k \) and \( J_k \), starting from an initial guess or heuristic. The objective is to minimize the approximation error, typically measured by the Frobenius norm, while maintaining low ranks \( r_k \) for computational efficiency.

This approach is scalable and particularly beneficial for high-dimensional tensors where direct manipulation is impractical. The method's key advantage is its scalability concerning tensor dimensionality, making it valuable for managing large, complex datasets.

\subsection{Quantization}
Quantization is a fundamental process in digital signal processing and numerical analysis, where a continuous spectrum of values is mapped to a finite set of discrete levels.
This process is necessary for encoding a probability distribution into a TT. A continuous distribution $p(x)$ can be approximated by a discrete probability vector $\mathbf{p}$, where
$$\mathbf{p} = [p(x_0), p(x_1),\dots, p(x_m)],$$
with the restriction that \(\sum_{i=0}^m p(x_i) = 1\). Throughout this section, we will assume that $m = 2^d-1$; however, this assumption is not strictly necessary.

These discrete points, \(\{x_i\}_{i=0}^m\), can then be encoded on $d=\lceil \log_2 (m+1) \rceil$ qubits with basis states $\ket{i_1 i_2 \dots i_d}$ where $i_k \in \{0, 1\}$. The encoding is given by
\begin{equation}
    \mathbf{P}\ket{0}^{\otimes_k d} = \sum_{i=0}^{2^d-1} \sqrt{p(x_i)} \ket{(i)_2};
\end{equation}

\noindent
here, $(i)_2$ denotes that the index $i$ is represented in base 2:
\begin{equation*}
    (i)_2 = (i_1, \ldots, i_{d}) = \sum_{k=1}^{d} i_k 2^{d-k}.
\end{equation*}

The process of encoding and learning matrices and vectors in a TT format is often referred to as the \emph{quantics tensor train} (QTT) \cite{OSELEDETS201070}. This technique allows us to express the vector \(\mathbf{p}\) of size \(2^d\) as a tensor 
\(A \in \mathbb{R}^{2 \times \cdots \times 2}\), such that 
\(p(x_i) = A(i_1, \ldots, i_d)\),
and apply TN methods to this representation to effectively encode it in TT format:
\begin{equation}
    \begin{tikzpicture}[
    node distance=2cm,
    label distance=-0.5cm,
    baseline=(current  bounding  box.center)]
    
    \node (A1) [circle, draw, minimum size=1.1cm] {\footnotesize{$A_{1}$}};
    \node (A2) [right of=A1, circle, draw, minimum size=1.1cm] {\footnotesize{$A_{2}$}};
    \node (A3) [right of=A2, circle, draw, minimum size=1.1cm] {\footnotesize{$A_{3}$}};
    \node (An1) [right of=A3, circle, draw, minimum size=1.1cm] {\footnotesize{$A_{d-1}$}};
    \node (An) [right of=An1, circle, draw, minimum size=1.1cm] {\footnotesize{$A_{d}$}};
    \node [below=0.3cm of A1] (x1) {$i_1$};
    \node [below=0.3cm of A2] (x2) {$i_2$};
    \node [below=0.3cm of A3] (x3) {$i_3$};
    \node [below=0.3cm of An1] (xn1) {$i_{d-1}$};
    \node [below=0.3cm of An] (xn) {$i_d$};
    \draw (A1) -- (A2);
    \draw (A2) -- (A3);
    \draw[dashed] (A3) -- (An1);
    \draw (An1) -- (An);
    \draw (A1) -- (x1);
    \draw (A2) -- (x2);
    \draw (A3) -- (x3);
    \draw (An1) -- (xn1);
    \draw (An) -- (xn);
    
    \end{tikzpicture}
\end{equation}

This technique can be extended to vectors and matrices of any dimension by decomposing the index ranges into prime factors. However, it is more straightforward to employ this method with dimensions that are powers of $2$. Such simplification proves particularly advantageous when representing TTs in quantum circuits.

The arrangement of tensor cores is crucial for the representative power of the TT, particularly when seeking low-rank representations with small bond dimensions. For one-dimensional distributions, the process is relatively straightforward: it involves determining the necessary precision or granularity and then creating a binary encoding of the domain points. However, for multivariate distributions, the process is more complex, requiring careful arrangement of the tensor cores to better represent the high-dimensional spaces.

In this study, we introduce and evaluate three specific quantization methods for encoding multivariate distributions into a TT: Sequential, Mirroring (particularly effective for 2D cases), and Interleaving. Each method offers a unique strategy for structuring the binary encoding to optimize the TT representation's effectiveness. We will briefly describe these three representation methods below.

\paragraph{Sequential}

The sequential approach to quantization involves encoding each dimension of the function in a step-by-step manner. For instance, for a function 
\(f : \mathbb{R}^2 \to \mathbb{R}\) with inputs  $(x,y)$, we shall approximate it by the following discrete function

\begin{equation}
    \mathbf{F}\ket{0}^{\otimes_k d} = \sum_{j,l=0}^{2^{\left(\frac{d}{2}\right)}-1} f(x_j, y_l) \ket{(j)_2(l)_2}
\end{equation}

The sequential encoding method can be represented as:

\begin{equation}
\begin{tikzpicture}[
  node distance=2cm,
  label distance=-0.5cm
]

\node (A1) [circle, draw, minimum size=1.4cm] {$A_{1}$};
\node (A2) [right of=A1, circle, draw, minimum size=1.4cm] {$A_{2}$};
\node (Ak) [right of=A2, circle, draw, minimum size=1.4cm] {$A_{\frac{d}{2}}$};
\node (B1) [right of=Ak, circle, draw, minimum size=1.4cm] {$B_{1}$};
\node (B2) [right of=B1, circle, draw, minimum size=1.4cm] {$B_{\frac{d}{2}-1}$};
\node (Bk) [right of=B2, circle, draw, minimum size=1.4cm] {$B_{\frac{d}{2}}$};

\node [below=0.3cm of A1] (x1) {$i_1$};
\node [below=0.3cm of A2] (x2) {$i_2$};
\node [below=0.3cm of Ak] (xn2) {$i_{\frac{d}{2}}$};
\node [below=0.3cm of B1] (xn2p1) {$i_{\frac{d}{2}+1}$};
\node [below=0.3cm of B2] (xn1) {$i_{d-1}$};
\node [below=0.3cm of Bk] (xn) {$i_d$};

\draw (A1) -- (A2);
\draw[dashed] (A2) -- (Ak);
\draw (Ak) -- (B1);
\draw[dashed] (B1) -- (B2);
\draw (B2) -- (Bk);

\draw (A1) -- (x1);
\draw (A2) -- (x2);
\draw (Ak) -- (xn2);
\draw (B1) -- (xn2p1);
\draw (B2) -- (xn1);
\draw (Bk) -- (xn);

\end{tikzpicture}
\end{equation}

\noindent
where A and B label the cores which represent the different indices of $j$ and $l$.
While this representation is efficient for univariate distributions, it fails to maintain the correlations between variables in higher dimensions, leading to poor convergence and unsuitable approximations for multivariate functions

\paragraph{Mirroring}

In the mirroring approach, the tensors are ordered such that the highest significant qubits of each variable are adjacent. This configuration adopts a mirrored sequencing pattern, designed to concentrate information within the structure of the tensor.

\begin{equation}
    \mathbf{F}\ket{0}^{\otimes_k d} = \sum_{j,l=0}^{2^{\left(\frac{d}{2}\right)}-1} f(x_j, y_l) \ket{\overline{(j)_2}(l)_2};
\end{equation}

\noindent
Here, $\overline{(j)_2}$ indicates that the bits in the binary representation are reversed, converting from big endian format, where the most significant bit is first to little endian format, where the least significant bit is first.

This approach is suitable for capturing the inter-variable correlations effectively while maintaining a structured tensor layout, when dealing with bivariate distributions, since it reduces the distance over which related information must propagate within the TT, potentially enhancing approximation accuracy.

However, there is not a straightforward manner in which this approach can be generalized for higher dimensions.

The mirroring encoding method can be represented as:
\begin{equation}
\begin{tikzpicture}[
  node distance=2cm,
  label distance=-0.5cm
]

\node (Ak) [circle, draw, minimum size=1.4cm] {$A_{\frac{d}{2}}$};
\node (A2) [right of=Ak, circle, draw, minimum size=1.4cm] {$A_{\frac{d}{2}-1}$};
\node (A1) [right of=A2, circle, draw, minimum size=1.4cm] {$A_{1}$};
\node (B1) [right of=A1, circle, draw, minimum size=1.4cm] {$B_{1}$};
\node (B2) [right of=B1, circle, draw, minimum size=1.4cm] {$B_{\frac{d}{2}-1}$};
\node (Bk) [right of=B2, circle, draw, minimum size=1.4cm] {$B_{\frac{d}{2}}$};

\node [below=0.3cm of Ak] (x1) {$i_1$};
\node [below=0.3cm of A2] (x2) {$i_2$};
\node [below=0.3cm of A1] (xn2) {$i_{\frac{d}{2}}$};
\node [below=0.3cm of B1] (xn2p1) {$i_{\frac{d}{2}+1}$};
\node [below=0.3cm of B2] (xn1) {$i_{d-1}$};
\node [below=0.3cm of Bk] (xn) {$i_d$};

\draw (Ak) -- (A2);
\draw[dashed] (A2) -- (A1);
\draw (A1) -- (B1);
\draw[dashed] (B1) -- (B2);
\draw (B2) -- (Bk);

\draw (Ak) -- (x1);
\draw (A2) -- (x2);
\draw (A1) -- (xn2);
\draw (B1) -- (xn2p1);
\draw (B2) -- (xn1);
\draw (Bk) -- (xn);

\end{tikzpicture}
\end{equation}

\paragraph{Interleaving}
The interleaving method maintains the original order of significance but alternates tensors between variables. This method interleaves the qubits of different variables, promoting inter-variable correlation. 

\begin{equation}
    \mathbf{F}\ket{0}^{\otimes_k d} = \sum_{j,l=0}^{2^{\left(\frac{d}{2}\right)}-1} f(x_j, y_l) \ket{(j)_2^{(1)}(l)_2^{(1)}(j)_2^{(2)}(l)_2^{(2)}\ldots(j)_2^{(\frac{d}{2})}(l)_2^{(\frac{d}{2})}};
\end{equation}

Our results show this is the most effective method for ensuring that all variables are represented throughout the tensor train, enhancing the overall fidelity of multivariate distributions.

The interleaving encoding method can be represented as:

\begin{equation}
\begin{tikzpicture}[
  node distance=2cm,
  label distance=-0.5cm
]

\node (A1) [circle, draw, minimum size=1.4cm] {$A_{1}$};
\node (B1) [right of=A1, circle, draw, minimum size=1.4cm] {$B_{1}$};
\node (A2) [right of=B1, circle, draw, minimum size=1.4cm] {$A_{2}$};
\node (B2) [right of=A2, circle, draw, minimum size=1.4cm] {$B_{2}$};
\node (Ak) [right of=B2, circle, draw, minimum size=1.4cm] {$A_{\frac{d}{2}}$};
\node (Bk) [right of=Ak, circle, draw, minimum size=1.4cm] {$B_{\frac{d}{2}}$};

\node [below=0.3cm of A1] (x1) {$i_1$};
\node [below=0.3cm of B1] (x2) {$i_2$};
\node [below=0.3cm of A2] (x3) {$i_3$};
\node [below=0.3cm of B2] (x4) {$i_4$};
\node [below=0.3cm of Ak] (xn1) {$i_{d-1}$};
\node [below=0.3cm of Bk] (xn) {$i_d$};

\draw (A1) -- (B1);
\draw (B1) -- (A2);
\draw (A2) -- (B2);
\draw[dashed] (B2) -- (Ak);
\draw (Ak) -- (Bk);

\draw (A1) -- (x1);
\draw (B1) -- (x2);
\draw (A2) -- (x3);
\draw (B2) -- (x4);
\draw (Ak) -- (xn1);
\draw (Bk) -- (xn);

\end{tikzpicture}
\end{equation}

\section{Mapping TT to Quantum Circuits}
\label{sec:tn_to_qc}
In this section, we introduce an algorithm for mapping a TT to a quantum circuit, building on the methodology described in \cite{Ran_2020}. 
Our goal is to explicitly construct a set of unitary operators, \(W_k\), representing quantum gates, that transform the initial state \(\ket{0}^{\otimes d}\) into a target quantum state
\begin{equation}
\ket{\psi_{tg}} = W_1 W_2 \ldots W_{d-1} W_d \ket{0}^{\otimes d},
\end{equation}
that satisfies
\begin{equation}\label{eq:target-mapping}
\braket{\psi_{tg}}{i_1, \ldots i_d} = \tfrac{1}{\mathcal{A}} \cdot A(i_1, \ldots, i_d),     
\end{equation}
\noindent
where $\mathcal{A}$ is a normalizing constant and \(A (i_1, \ldots, i_d) \in \mathbb{R}^{2^d}\) is the $d$ dimensional TT that encodes the target distribution, cf. Section~\ref{sec:tn} for more details.

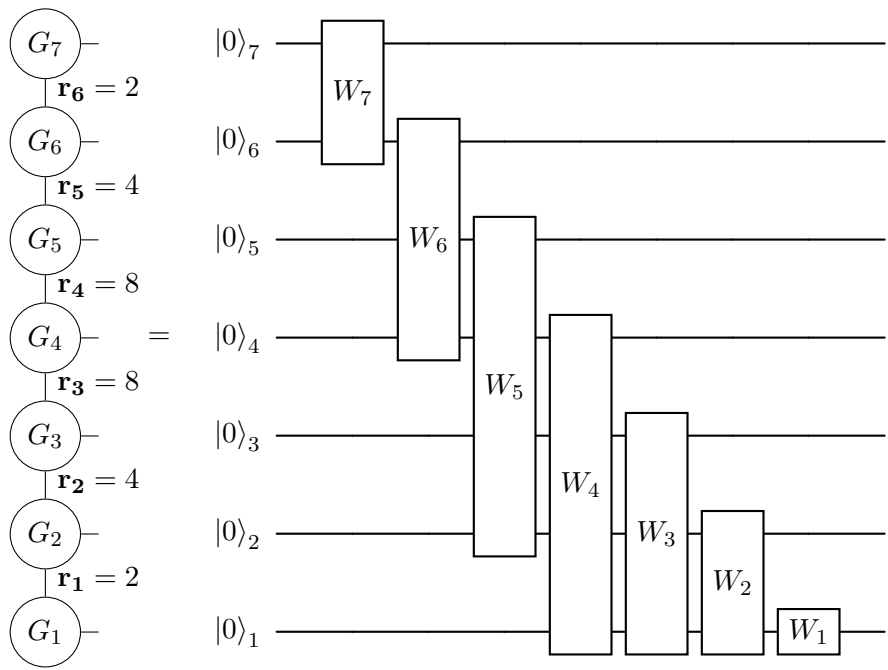
\begin{figure}[h!]
\centering
\begin{tikzpicture}
    \node[draw, circle] (G1) at (0,-3.9) {$G_{1}$};
    \node[draw, circle] (G2) at (0,-2.6) {$G_{2}$};
    \node[draw, circle] (G3) at (0,-1.3) {$G_{3}$};
    \node[draw, circle] (G4) at (0,0) {$G_{4}$};
    \node[draw, circle] (G5) at (0,1.3) {$G_{5}$};
    \node[draw, circle] (G6) at (0,2.6) {$G_{6}$};
    \node[draw, circle] (G7) at (0,3.9) {$G_{7}$};
    
    \draw (G1) -- (G2) -- (G3) -- (G4) -- (G5) -- (G6) -- (G7);
    \draw (G1) -- ++(0.7,0);
    \draw (G2) -- ++(0.7,0);
    \draw (G3) -- ++(0.7,0);
    \draw (G4) -- ++(0.7,0);
    \draw (G5) -- ++(0.7,0);
    \draw (G6) -- ++(0.7,0);
    \draw (G7) -- ++(0.7,0);

    \node (B1) at (0.7,-3.2) {$\mathbf{r_1} = 2$};
    \node (B2) at (0.7,-1.9) {$\mathbf{r_2} = 4$};
    \node (B3) at (0.7,-0.6) {$\mathbf{r_3} = 8$};
    \node (B4) at (0.7,0.7) {$\mathbf{r_4} = 8$};
    \node (B5) at (0.7,2) {$\mathbf{r_5} = 4$};
    \node (B6) at (0.7,3.3) {$\mathbf{r_6} = 2$};
    
    \node at (1.5, 0) {$=$};
\end{tikzpicture}
\hspace{0.01cm}
\begin{quantikz}[wire types={q,q,q,q,q,q,q}, classical gap = 0.03cm, row sep = {1.3cm,between origins},column sep = {1cm,between origins}, align equals at = 7.29]
\lstick[label style={inner sep=5pt}]{$\ket{0}_7$}&\gate[2]{W_7}&&&&&&& \\
\lstick[label style={inner sep=5pt}]{$\ket{0}_6$}&&\gate[3]{W_6}&&&&&& \\
\lstick[label style={inner sep=5pt}]{$\ket{0}_5$}&&&\gate[4]{W_5}&&&&& \\
\lstick[label style={inner sep=5pt}]{$\ket{0}_4$}&&&&\gate[4]{W_4}&&&& \\
\lstick[label style={inner sep=5pt}]{$\ket{0}_3$}&&&&&\gate[3]{W_3}&&& \\
\lstick[label style={inner sep=5pt}]{$\ket{0}_2$}&&&&&&\gate[2]{W_2}&& \\
\lstick[label style={inner sep=5pt}]{$\ket{0}_1$}&&&&&&&\gate{W_1}& \\
\end{quantikz}

\caption{Equivalence between a TT and a quantum circuit. The circuit illustrates the step-by-step application of unitary operations on seven initially unentangled qubits (each in the \( \ket{0} \) state). The sequence of operations begins with the qubit associated with the right-most, or last, core and moves leftward. Each unitary operator is applied to entangle its corresponding qubit with the rest of the system, following a specific pattern that aligns with the tensor train (TT) configuration.
}
\label{fig:tt-to-qc}
\end{figure}

Our constructions leverages the explicit decomposition of \(A (i_1, \ldots, i_d) \in \mathbb{R}^{2^d}\) into TT-cores: 
\begin{equation}\label{eq:tt-format}
   A(i_1, \ldots, i_d) = \sum_{r_0, r_1, \ldots, r_d} G_1(r_0, i_1, r_1) G_2(r_1, i_2, r_2) \cdots G_d(r_{d-1}, i_d, r_d),
\end{equation}
\noindent
for building the unitary operators \(W_k\). 
Starting from $k=1$ we will build the circuit from the final core to the first one, as illustrated in Figure \ref{fig:tt-to-qc} below. We start by reshaping the TT-core \(G_1(r_{0}, i_1, r_1)\) into a \(\mathbb{C}^{2 \times n_1} \) matrix\footnote{In this specific case, range of the index $r_0$ is one making the reshaping trivial.}
\(
M_{1}(i_1, r_1) = G_1(0, i_1, r_1)
\),
and factor it (SVD theorem) as \(M_{1} = U \Sigma V^{*}\), where \(U \in \mathbb{C}^{2 \times 2} \) and \(V^{*} \in \mathbb{C}^{n_1 \times n_1} \) are unitary matrices and \(\Sigma \in \mathbb{C}^{2 \times n_1} \) is a diagonal matrix containing the singular values of $M_1$. 
Substituting this into the right-hand side of \eqref{eq:tt-format} gives
\begin{equation}\label{eq:iteration-split}
    \sum_{j=0, 1} U (i_1, j)  \left[ \sum_{r_2, \ldots, r_d} \tilde{G}_2(j, i_2, r_2) \cdots G_d(r_{d-1}, i_d, r_d) \right], 
\end{equation}
where 
\(
\tilde{G}_2(j, i_2, r_2) = \sum_{r_1} \big(\Sigma V^{*}\big)(j, r_1) G_2(r_1, i_2, r_2)
\). With a slight abuse of notation we shall drop the upper tilde and denote $\tilde{G}_2 = G_2$.

The first sum in \eqref{eq:iteration-split} can be seen as the action of the (unitary) operator on \(\mathbb{C}^2 \otimes \mathbb{C}^{2^{d-1}}\): 
\[
\begin{split}
\ket{0} \otimes \ket{v}& \mapsto U(0,0) \ket{0} \otimes \ket{v} + U(1,0) \ket{1} \otimes \ket{v}; \\
\ket{1} \otimes \ket{v}& \mapsto U(0,1) \ket{0} \otimes \ket{v} + U(1,1) \ket{1} \otimes \ket{v}.
\end{split}
\]
We will set $W_1$ to be such an operator. Next, we turn out attention to the term within the brackets in \eqref{eq:iteration-split}. First, we notice that the range of the index $j$ can be reduced by truncating singular value matrices \(\Sigma\): the rows  \(\Sigma(j, :)\) are null if $j \geq 2 \wedge n_1$. In particular we may see the term within brackets as a new tensor:
\[
A^{(2)}(j, i_2, \ldots, i_d) = \sum_{r_2, \ldots, r_d} G_2(j, i_2, r_2) \cdots G_d(r_{d-1}, i_d, r_d),
\]
where the index $j$ ranges from $0$ to $2 \wedge n_1-1$. At this point it will be useful to make the assumption that the bond dimensions $n_k = 2^{l_k}$, so that \(2 \wedge n_1 = 2^{1 \wedge l_1}\),  this will facilitate the mapping between TT indices and qubits and make the approach and the construction of the quantum gates more straightforward. Practically, we can always impose this condition to the TT by augmenting the virtual dimensions when necessary.

To iterate the above procedure, we start with 
\[
A^{(k)}(j, i_k, \ldots, i_d) = \sum_{r_2, \ldots, r_d} G_k(j, i_k, r_k) \cdots G_d(r_{d-1}, i_d, r_d),
\]
where $j=0, \ldots, 2^{(k-1) \wedge l_{k-1}} - 1$. We reshape the TT-core $G_k$ into a matrix 
\[
M_{k} \in \mathbb{C}^{m_k \times n_k}, \quad 
\text{
where $m_k = 2 \big( 2^{(k-1) \wedge l_{k-1}}\big) = 2^{k \wedge (1+ l_{k-1})}$.
}
\]
Next, we factor \(M_{k} = U \Sigma V^* \) using singular values and obtain the $k$th unitary operator $W_k = U$ on the circuit, which acts on the qubits $k, \ldots , k - k \wedge l_{k-1}$. The terms $\Sigma V^* $ are passed to $G_{k + 1}$ which becomes a new TT-core whose index $j$ ranges from $0$ to $2^{k \wedge l_{k-1} + 1} \wedge 2^{l_k}$. Finally we notice that by construction of the TT-cores $n_{k} \leq 2 n_{k-1}$, so that,  $2^{k \wedge l_{k-1} + 1} \wedge 2^{l_k} =  2^{k \wedge l_{k} }$ allowing us to iterate the algorithm. Notice that when reshaping the final TT-core $ G_d(j, i_d, r_d)$ the matrix $M_d \in \mathbb{C}^{2 \times 1} $ is a column matrix. In particular, it has a unique singular value $\mathcal{A}$, which stands for the normalization constant in \eqref{eq:target-mapping}. 

We summarize this procedure in Algorithm~\ref{alg:tn_to_qc}.

\begin{algorithm}
  \caption{TT to Quantum Circuit}\label{alg:tn_to_qc}
  \begin{algorithmic}
    \State \textbf{Input} $G=[G_1, \ldots, G_d]$ \Comment{List of TT-cores}
    \State \textbf{Output} $W=[W_1, \ldots, W_d]$ \Comment{List unitary operators}
    \State $W \gets [ \, ]$

    \For{$i \gets 1$ to $\mathrm{length}(G)$}  
        \State $M \gets \text{reshape}(G[i])$
        \State $U, \Sigma, V^* \gets \mathbf{SVD} (M)$ 
        \State $\Sigma \gets \mathbf{truncate}(\Sigma)$
        \State $R \gets (\Sigma V^*) \otimes I_2$
        \State $G[i+1] \gets R G[i+1] $ 
        \State $W.\text{insert}(G[i])$
    \EndFor
    \State \Return {$W$}
  \end{algorithmic}
\end{algorithm}

\subsection{Reducing circuit depth}

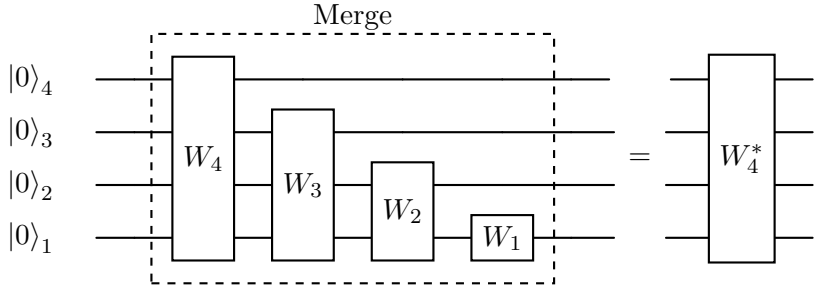
\begin{figure}[htb!]
\centering
\begin{quantikz}[wire types={q,q,q,q}, classical gap = 0.08cm, row sep = {0.7cm,between origins}]
\lstick[label style={inner sep=15pt}]{$\ket{0}_4$}&&\gate[4]{W_4}\gategroup[4,steps=4,style=dashed]{Merge}&&&&&\midstick[4,brackets=none]{=} &\gate[4]{W_4^*}& \\
\lstick[label style={inner sep=15pt}]{$\ket{0}_3$}&&&\gate[3]{W_3}&&& &&& \\
\lstick[label style={inner sep=15pt}]{$\ket{0}_2$}&&&&\gate[2]{W_2}&& &&& \\
\lstick[label style={inner sep=15pt}]{$\ket{0}_1$}&&&&&\gate{W_1}& &&&
\end{quantikz}
\caption{``Merging'' unitary operations to reduce the depth of the quantum circuit. The dashed box encloses a subsection of the original circuit that can be replaced by a single, more complex unitary operation \( W^*_4 \). This operation has the same overall effect as the sequence of operations \( W_4, W_3, W_2, W_1 \) on the last four qubits.
}
\label{fig:merging}
\end{figure}

The depth of a quantum circuit can be reduced by merging smaller unitary gates with preceding larger ones that affect the same qubits. This method, illustrated in Figure \ref{fig:merging}, results in a new unitary gate of the same size as the preceding larger one, while performing all the transformations of the merged resulting from the merged gates. The final circuit is equivalent to the one in Figure \ref{fig:tt-to-qc} but with a reduced circuit depth. Reducing the depth simplifies the quantum circuit by decreasing the number of sequential operations, which can shorten execution times and reduce the likelihood of quantum decoherence. This enhances the reliability and efficiency of the circuit on quantum hardware.

\subsection{Implementation in Qiskit}
To test, and more easily assess this method for encoding data into Quantum Circuits, the described process was implemented in Python, and subsequently the circuits were created using Qiskit version 0.45.0.

First the TT-cross algorithm described in \ref{subsec:tt-cross} is used to construct the desired TT. Then, after applying Algorithm~\ref{alg:tn_to_qc} to the resulting TT-cores, the circuit was implemented using the Qiskit UnitaryGate method. The results from these experiments can be found in Section~\ref{sec:res}.

\section{Results}
\label{sec:res}
This section presents the results from experiments conducted with both univariate and multivariate distributions provided by Ita\'u Unibanco, specifically used for pricing financial instruments in practical applications. Our focus was primarily on log-normal distributions, with parameters calibrated using real market data.

We evaluated the training process as a function of number of qubits, which determines the discretization precision of the distribution, analyzed the scalability of the encoding method, and employed different KPIs to assess the distributions generated by the circuit executions. Our findings highlight the
scalability of the TT-cross method, which significantly outperforms standard Qiskit methods. This demonstrates a substantial advantage as circuits scale up, making the TT-cross method a promising approach for practical implementations.

\begin{description}
    \item[Kolmogorov-Simrnov distance] The Kolmogorov-Smirnov (KS) distance is a non-parametric measure used to quantify the maximum difference between the cumulative distribution functions (CDFs) of two probability distributions. Mathematically, given two cumulative distribution functions $F(x)$ and $G(x)$, the KS distance, denoted by $D$, is defined as:
    $$D = \sup_x | F(x) - G(x) |,$$
    where \( \sup_x \) denotes the supremum over all values of \( x \). The KS distance measures the greatest vertical distance between the two CDFs at any point on the real line.
\end{description}

\begin{figure}[htb!]
    \centering
    \begin{minipage}{0.5\textwidth}
        \centering
        \includegraphics[width=0.9\textwidth]{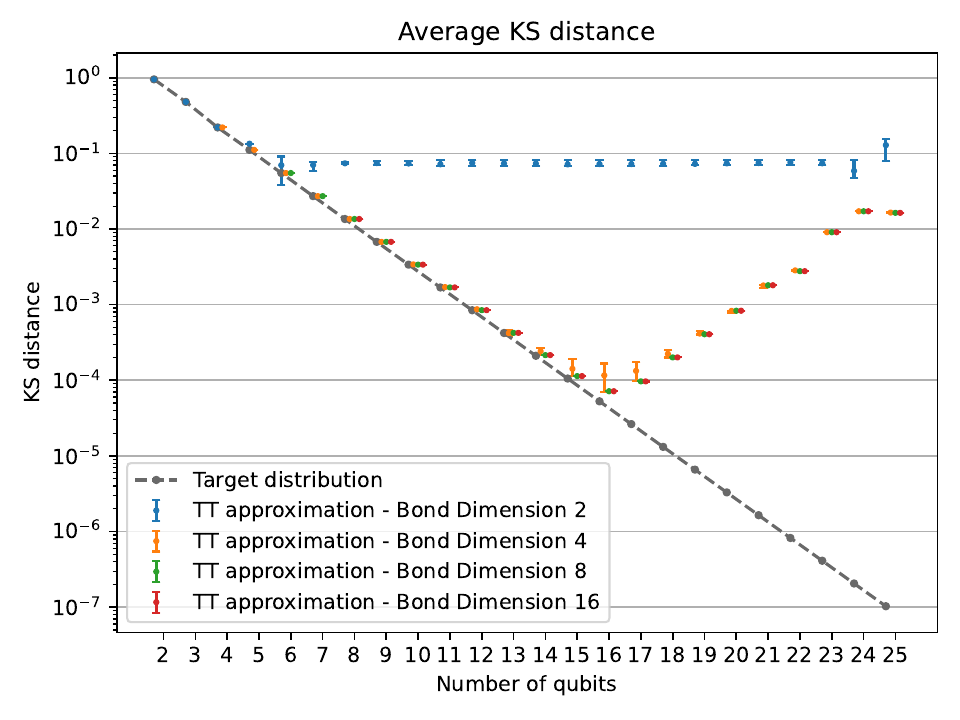}
    \end{minipage}%
    \begin{minipage}{0.5\textwidth}
        \centering
        \includegraphics[width=0.9\textwidth]{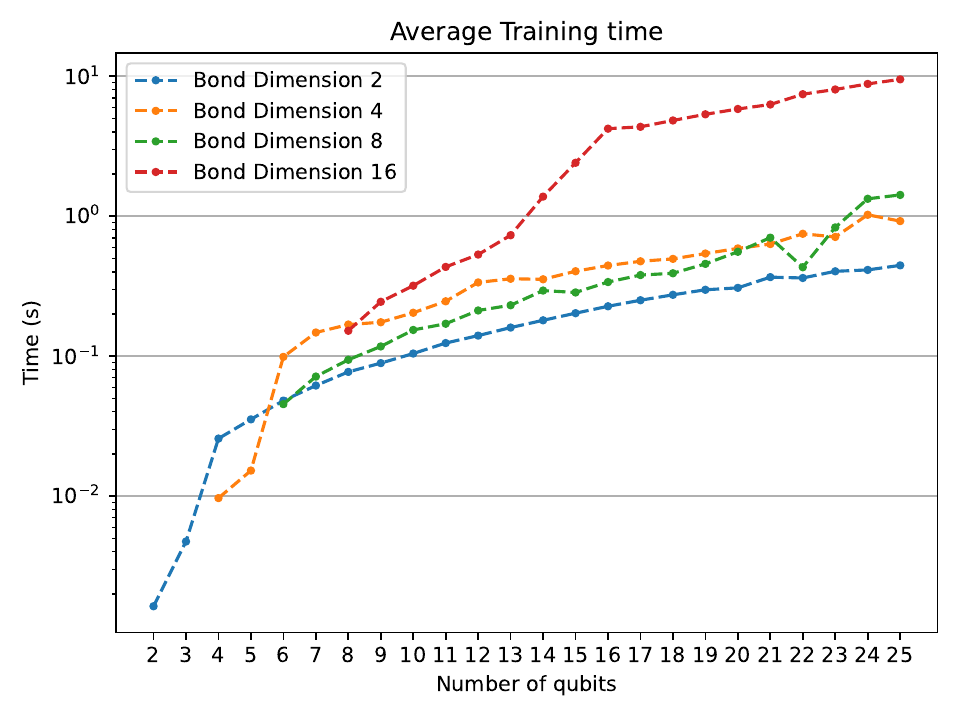}
    \end{minipage}
    
    \caption{Metrics obtained for approximating a univariate Log-Normal distribution: on the left, the KS distance obtained as a function of the number of qubits; on the right, the average training time, also as a function of the number of qubits.}
    \label{fig:avg_lognormal}
\end{figure}

We begin by presenting the results of the univariate case. In Figure~\ref{fig:avg_lognormal}, the KS distance is used to illustrate the TT-cross approximation for the univariate log-normal distribution. Notably, the KS distance between the obtained and original distributions achieves a precision of $7.2 \times 10^{-5}$ with 16 qubits (65,536 discretization points). However, beyond this point, the precision of the approximation decreases, even when increasing the bond dimension beyond 8. This decrease in precision can be attributed to the accumulation of small errors across the domain. While the TT-cross method maintains a low error at individual points, the overall error accumulates as the domain size grows, leading to a reduced accuracy in the KS distance.Additionally, although theoretically a larger bond dimension should enhance precision, in practice, bond dimensions larger than 8 lead to convergence issues when the number of qubits exceeds 15. This explains why further increasing the bond dimension does not yield improved results in the approximation.
Furthermore, the TT-cross algorithm shows desirable behavior with respect to training. As indicated in Figure~\ref{fig:avg_lognormal}, the training time scales logarithmically with the number of points.

\begin{figure}[htb!]
    \centering
    \begin{minipage}{0.5\textwidth}
        \centering
        \includegraphics[width=0.9\textwidth]{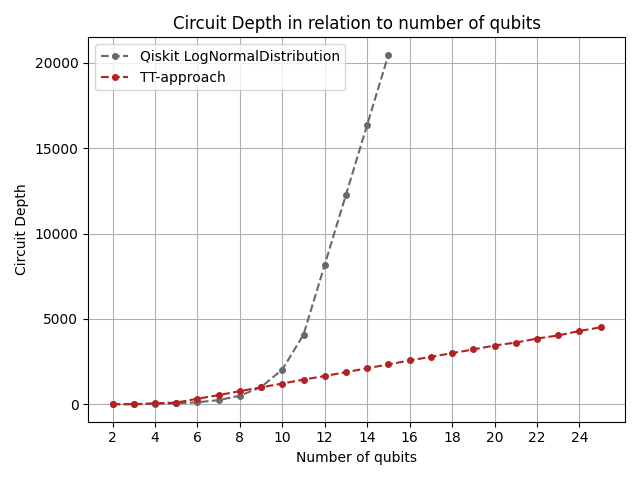}
    \end{minipage}%
    \begin{minipage}{0.5\textwidth}
        \centering
        \includegraphics[width=0.9\textwidth]{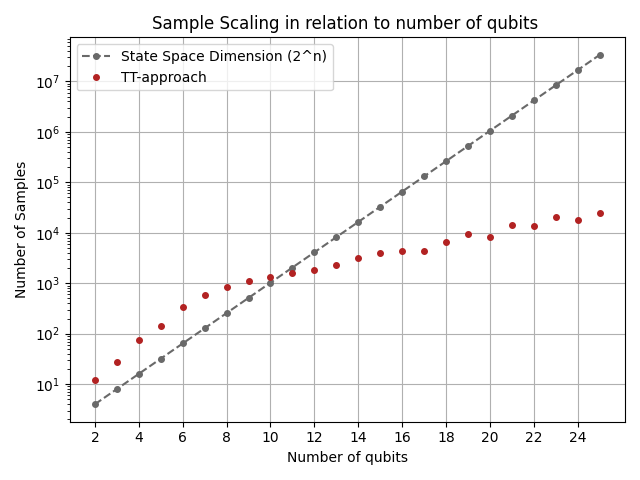}
    \end{minipage}
    
    \caption{The left-hand plot illustrates circuit depth as a function of the number of qubits, comparing Qiskit's built-in implementation of the LogNormal distribution with the TT-approach, with the bond dimesnion capped at 8. It demonstrates that the TT-method achieves significantly better scalability in terms of circuit depth as the number of qubits increases. The right-hand plot assesses training complexity, comparing the computational effort required for Qiskit's built-in implementation and the TT-method. We observe that the TT-approach exhibits logarithmic scaling with respect to the number of discretization points, in contrast to the linear scaling of the Qiskit approach.}
    \label{fig:samples-circuit_depth}
    
\end{figure}

Figure~\ref{fig:samples-circuit_depth} highlights the primary advantages of the TT-method: a significantly reduced circuit depth and fewer operations required for training. The left-hand plot illustrates circuit depth as a function of the number of qubits, comparing the TT-method with Qiskit's built-in implementation. The Qiskit method exhibits a clear exponential increase in circuit depth, making it impractical for large numbers of qubits. In contrast, the TT-method demonstrates linear scaling, making it a better option for implementing large circuits. The right-hand plot compares the computational effort required for both methods. The Qiskit method, based on the Grover-Rudolph technique, scales linearly with the number of discretization points, resulting in exponential scaling with the number of qubits. Conversely, the TT-approach shows logarithmic scaling with respect to the number of points, offering a significant advantage.

\begin{figure}[htb!]
    \centering
\includegraphics[width=0.8\textwidth]{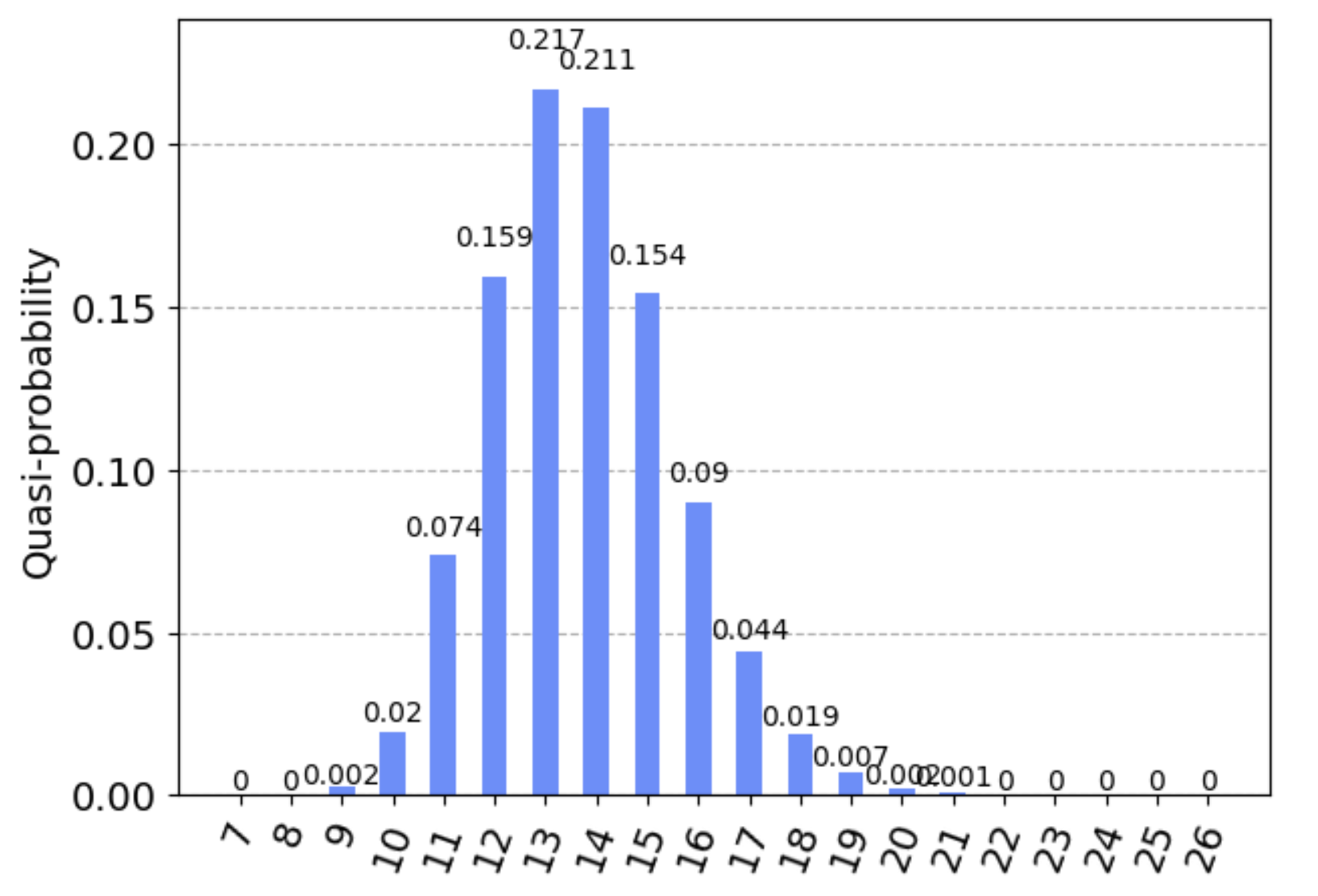}
    \caption{Empirical distribution of the log-normal distribution encoded using the TT-cross method with 5-qubit precision. This distribution was obtained by executing and measuring the quantum circuit, constructed through the TT-cross method described earlier, on a simulator. The circuit was run 10,000 times to ensure statistical significance, resulting in an accurate and reliable representation of the expected outcomes.}
    \label{fig:simulator}
\end{figure}

To complete our analysis of the univariate case, we implemented the algorithm on actual quantum hardware. For this experiment, we used IBM's Eagle R3 processors \cite{IBMeagle}. We evaluated our implementation on a 5-qubit system to assess the hardware's performance and the effect of quantum noise on the encoded distribution\footnote{We selected a 5-qubit system for its balance between computational feasibility and representation accuracy. This configuration is large enough to capture the essential shape of a log-normal distribution, providing a meaningful depiction of its properties. At the same time, the circuit depth for this setup remains relatively moderate, typically involving one hundred gates before transpiling.}. The expected results for this experiment are shown in Figure~\ref{fig:simulator}, which depicts the target empirical distribution obtained from a quantum simulator without noise. 

\begin{figure}[htb!]
    \centering
    \begin{minipage}{0.5\textwidth}
        \centering
        \includegraphics[width=0.95\textwidth]{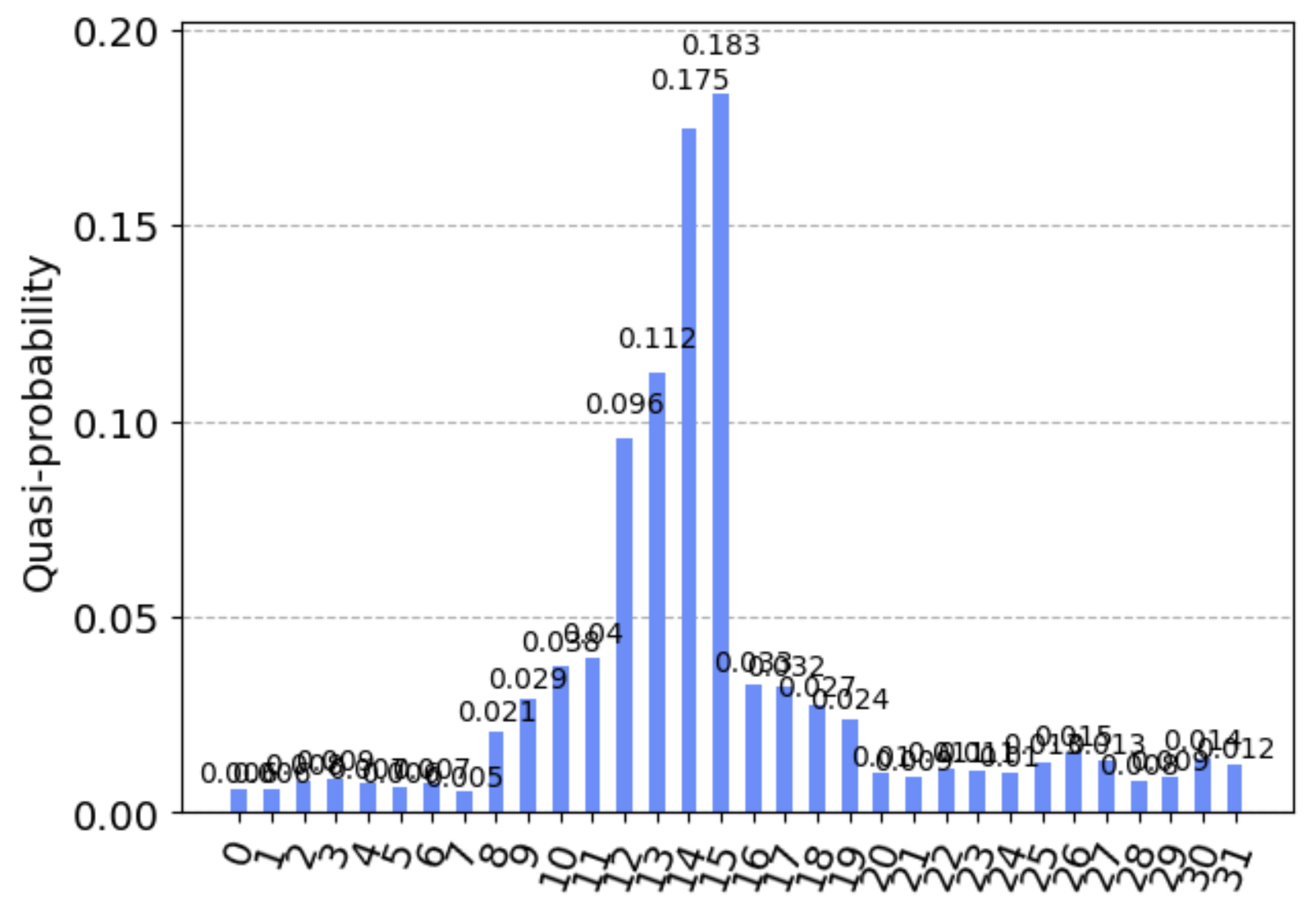}
    \end{minipage}%
    \begin{minipage}{0.5\textwidth}
        \centering
        \includegraphics[width=0.95\textwidth]{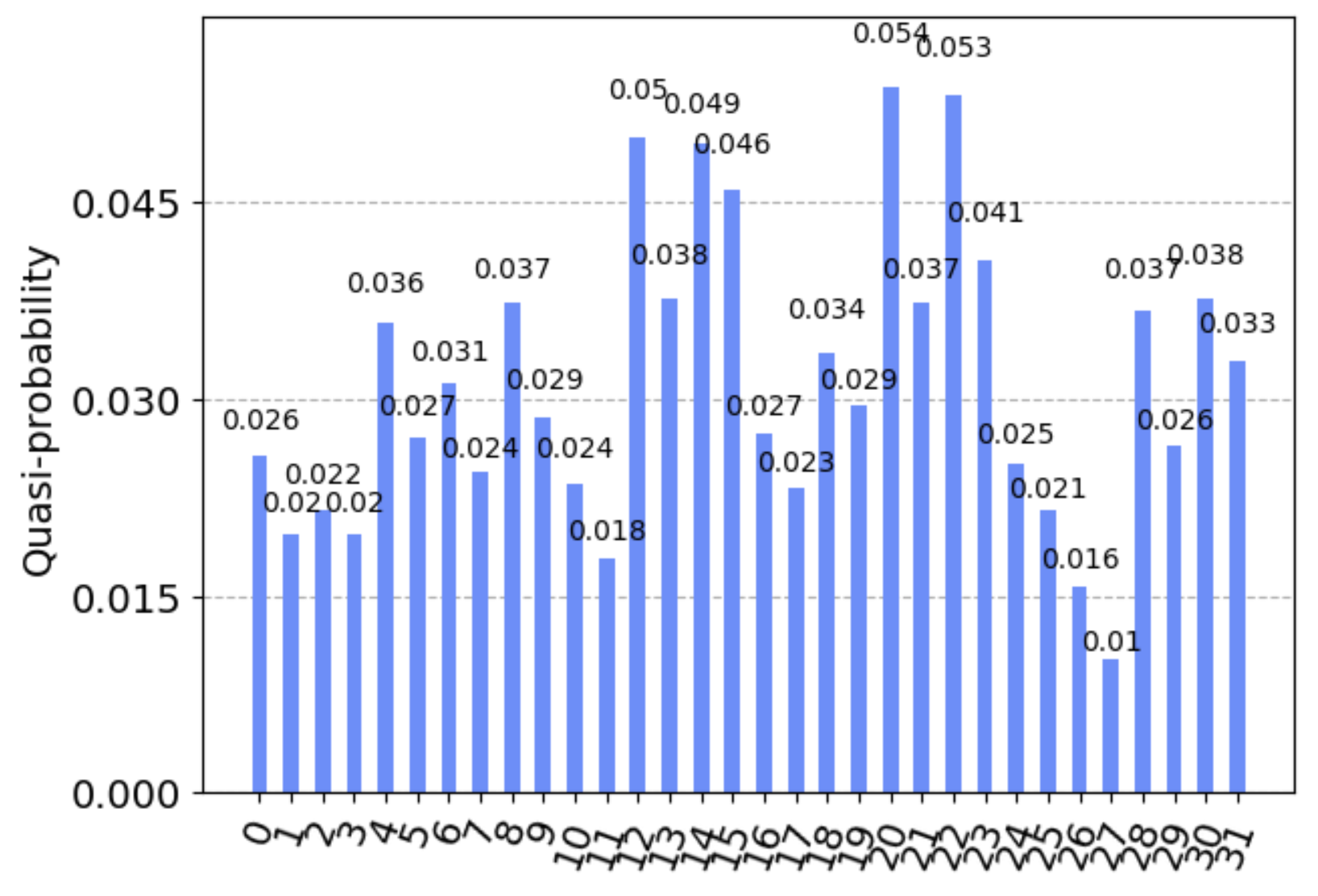}
    \end{minipage}
    
    \caption{Empirical results of encoding a log-normal distribution in 5 qubits using the TT-cross method on the IBM Eagle R3 machine, tested under different optimization levels. On the left, with optimization level set to 1, the distribution does not perfectly match the expected log-normal shape but retains recognizable patterns and captures the overall structure of the distribution. On the right, with optimization level set to 3, the distribution appears more uniform, indicating that the noise levels in the system hinder an accurate representation of the log-normal distribution.}
    \label{fig:ibm-experiment}
    
\end{figure}

The results obtained from the IBM processor are shown in Figure~\ref{fig:ibm-experiment}. To examine the impact of compiler optimizations on the execution outcomes, we conducted experiments under two distinct optimization settings. On one hand, the left-hand plot indicates that some patterns of the expected distribution (Figure~\ref{fig:simulator}) are preserved. On the other hand, when the optimization level is set to 3, as shown in the right-hand plot, the outcome is significantly worse. This highlights the susceptibility of quantum machines to noise and underscores the importance of optimizing circuits and implementing the appropriate error correction techniques to improve performance.

In the remainder of this section, we present the results obtained for multivariate distributions. Our analysis included distributions with up to five dimensions (representing five assets), each represented by up to ten qubits.
This approach allowed us to use up to $2^{50}$ discretization points to approximate the target distributions. Such extensive reach demonstrates the exceptional capacity of tensor networks to compress large amounts of information (data) while using relatively minimal computational resources. This efficiency in managing large-scale data highlights the potential of tensor networks for complex data representation and computation.

\begin{description}
    \item[Kullback-Leibler (KL) Divergence]
    The Kullback-Leibler (KL) divergence, also called the relative entropy, is a fundamental concept in information theory that measures the difference between two probability distributions. Formally, for two probability distributions $P(x)$ and $Q(x)$ defined over the same probability space, where $P(x)$ is typically considered the true distribution and $Q(x)$ the approximating distribution, the KL divergence from $Q$ to $P$ is defined as:
    
    $$
    D_{KL}(P \parallel Q) = \sum_{x} P(x) \log \left(\frac{P(x)}{Q(x)}\right)
    $$
    
    The KL divergence is always non-negative and equals zero if and only if $P(x) = Q(x)$
    for all $x$. It is not symmetric, meaning that $D_{KL}(P \parallel Q) \neq D_{KL}(Q \parallel P)$, and therefore does not constitute a true metric.
\end{description}

\begin{figure}[htb!]
    \centering
    \begin{minipage}{0.5\textwidth}
        \centering
        \includegraphics[width=0.9\textwidth]{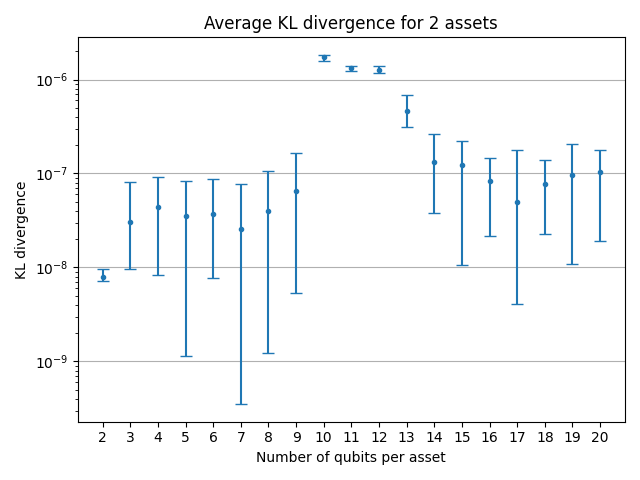}
    \end{minipage}%
    \begin{minipage}{0.5\textwidth}
        \centering
        \includegraphics[width=0.9\textwidth]{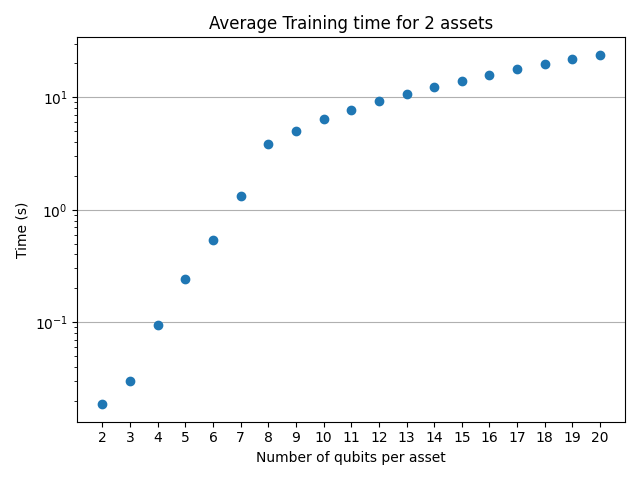}
    \end{minipage}
    
    \caption{Metrics obtained for approximating a bivariate Log-Normal distribution. The left plot shows the KL divergence as a function of the number of qubits per dimension, while the right plot illustrates the average training time, also as a function of the number of qubits per dimension.}
    \label{fig:lognormal_mirror}
\end{figure}

In Figure~\ref{fig:lognormal_mirror}, we present the results for the two-dimensional case. This is distinct from higher-dimensional cases because the 2-D representation allows us to use mirroring quantization, which we found led to better outcomes. We measure the accuracy of the represented distribution using relative entropy, also known as Kullback–Leibler (KL) divergence. The lower KL divergence (shown on the left-hand side of the plot) indicates a closer match between the encoded distribution and the target discrete log-normal distribution, highlighting the effectiveness of the quantization techniques employed. The right-hand plot also demonstrates how the training time scales well with the number of qubits, indicating that the method remains feasible even with high discretization levels.

\begin{figure}[htb!]
    \centering
    \begin{minipage}{0.5\textwidth}
        \centering
        \includegraphics[width=0.9\textwidth]{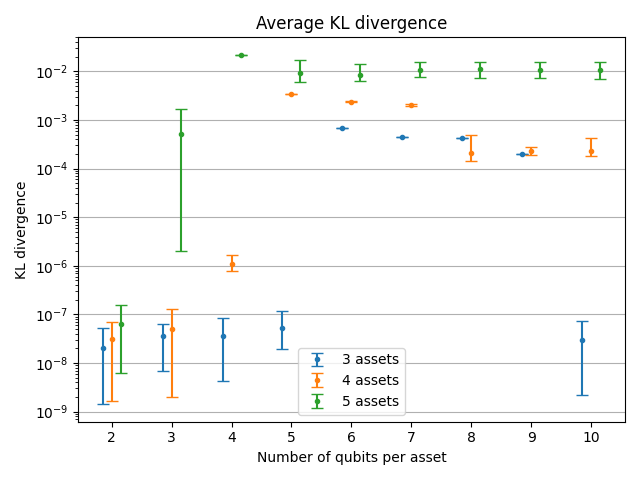}
    \end{minipage}%
    \begin{minipage}{0.5\textwidth}
        \centering
        \includegraphics[width=0.9\textwidth]{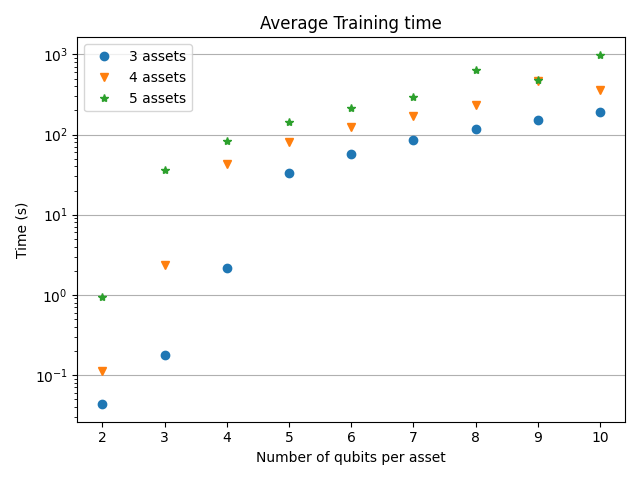}
    \end{minipage}
    
    \caption{KPIs for approximating multivariate log-normal distributions up to 5 dimensions. The left-hand plot illustrates KL divergence as a function of the number of qubits per dimension. The right-hand plot shows the average training time, also as a function of the number of qubits per dimension.}
    \label{fig:lognormal_inter}
\end{figure}

The results for higher dimensions are depicted in Figure~\ref{fig:lognormal_inter}. We used the interleaving quantization technique, which we found yielded the best outcome. Although the KL divergence is higher than in the 2D case, it remains relatively low ($\sim 10^{-2}$ in the worst-case scenario), indicating that the TT method can effectively approximate distributions in high dimensions. Additionally, the training time exhibits a logarithmic pattern, demonstrating that the algorithm can efficiently handle high-dimensional data.

\begin{figure}[htb!]
    \centering
    \begin{minipage}{0.5\textwidth}
        \centering
        \includegraphics[width=0.9\textwidth]{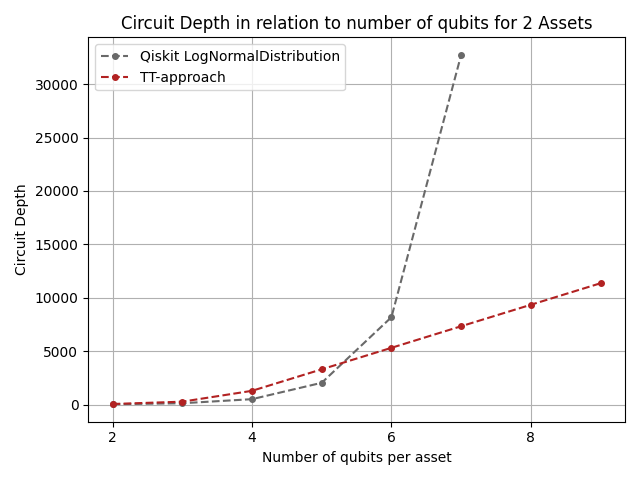}
    \end{minipage}%
    \begin{minipage}{0.5\textwidth}
        \centering
        \includegraphics[width=0.9\textwidth]{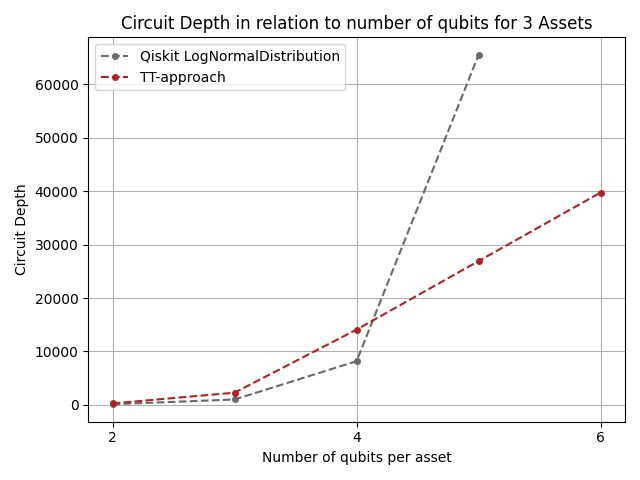}
    \end{minipage}
    
    \caption{Circuit depth as a function of the number of qubits per dimension, comparing Qiskit’s built-in implementation of the log-normal distribution with the TT approach. The left plot shows values for a bivariate distribution, while the right plot shows values for a 3D distribution. Similar to the 1D case, these plots demonstrate that the TT method achieves better scalability in terms of circuit depth as the number of qubits increases.}
    \label{fig:samples-circuit_depth-2d}
    
\end{figure}

Finally, in Figure~\ref{fig:samples-circuit_depth-2d}, we illustrate the circuit depth scaling of the TT-method. For clarity, we only present the 2D and 3D cases. The circuit depth achieved by the TT approach exhibits a linear scaling behavior in relation to the number of qubits, significantly outperforming Qiskit’s built-in method for large systems. This demonstrates the TT-method's superior efficiency and scalability compared to traditional techniques. A logarithmic scaling behavior is observed with the complexity, in relation to the number of points, compared to a linear scaling from Qiskit’s built-in method, further highlighting the advantages of the TT approach.

Given the high number of qubits required to represent these distributions and the insights gained from the univariate case, we concluded that conducting tests on actual quantum hardware would not yield meaningful results and could misleadingly suggest that the TN method is inappropriate. Instead, we focused our efforts on simulations, which provide a more controlled environment for evaluating the theoretical performance and capabilities of tensor networks in encoding and processing complex multivariate distributions. This approach ensures that our findings are robust and accurately reflect the true potential of the methods employed, without the confounding influence of hardware-induced noise. By using simulations, we can isolate the effects of our algorithms and techniques, allowing for a clearer understanding of their effectiveness and limitations.

\section{Conclusions}
\label{sec:conclusion}

This project explored the application of QMC techniques to financial modeling, addressing the critical challenge of encoding probability distributions into quantum computers -- a major bottleneck for QMC efficiency. By focusing on the financial distributions used by Ita\'u Unibanco for pricing financial instruments, we aim to promote the integration of quantum computing in the financial sector, thereby improving financial modeling and pricing strategies.

The TT-cross method was shown to be particularly promising, providing a representation that is both efficient in approximating general probability distributions and expressible as a quantum circuit. The approach involved approximating distributions using the TT-cross method to generate a TT representation. This representation captures the essence of the probability distribution in a format suitable for quantum computation. State-of-the-art algorithms were then used to map the TT onto a quantum circuit, transforming the tensor network into a sequence of quantum operations/gates applied to a quantum register. This mapping allows for the TT structure to be efficiently captured within quantum hardware.

A benchmark was carried out to evaluate the TT-cross approximation and its mapping to quantum circuits. The traditional Grover-Rudolph method for state preparation, commonly used in QMC, becomes impractical for large qubit systems due to its scaling complexity. In contrast, the TT-cross approach mitigates the curse of dimensionality, with circuit and operation complexity scaling logarithmically with system size. This method significantly improves existing probability loading techniques, resulting in more compact and optimized circuits well-suited for near-term quantum machines. 
However, fully implementing this on quantum computers will require significant advancements in both hardware technology and algorithms, particularly in error mitigation techniques. Current quantum hardware is still prone to various types of noise and errors that can affect the accuracy and reliability of computations. 

Future research directions could explore other methods for approximating distributions in a tensor-train format, as the TT-cross algorithm is not the only approach available \cite{doi:10.1137/15M1036919}\cite{GORODETSKY20181219}\cite{GORODETSKY201959}. Further development of alternative methods could lead to improvements in accuracy and efficiency. Additionally, some distributions have well-known analytical formulas,  which could potentially be used to directly encode them in a tensor-train format. This approach, if feasible, could eliminate the need for approximation through algorithms like TT-cross and thereby avoiding approximation errors and algorithmic limitations. Although this line of research falls outside the scope of the current project, it presents a promising area for further investigation, with potential to enhance the efficiency and precision of quantum computation in various fields.

\hfill \break

\noindent
\textbf{Disclaimer.}
This paper is a research collaboration between Ita\'u Unibanco and Multiverse Computing. Any opinions, findings, conclusions or recommendations expressed in this material are those of the authors and do
not necessarily reflect the views of Ita\'u Unibanco. This paper is not
and does not constitute or intend to constitute investment advice or any
investment service. It is not and should not be deemed to be an offer
to purchase or sell, or a solicitation of an offer to purchase or sell, or
a recommendation to purchase or sell any securities or other financial
instruments. Moreover, all data used in this study is compliant with the
Brazilian General Data Protection Law.

\bibliographystyle{abbrv}
\bibliography{references}

\end{document}